\documentclass[twocolumn,preprintnumbers,amsmath,nofootinbib,amssymb]{revtex4-1}

\usepackage{amsmath,amssymb,epsfig,graphicx}

\usepackage{dcolumn}

\def\be{\begin{eqnarray}}
\def\ee{\end{eqnarray}}
\def\bea{\begin{eqnarray}}
\def\eea{\end{eqnarray}}

\begin{document}

\preprint{}

\title{
${\cal N}=1$ conformal dualities
}

\author{
Shlomo S. Razamat$^1$ and Gabi Zafrir$^2$
}

\

\affiliation{
$^1$Department of Physics, Technion, Haifa 32000, Israel\\ 
$^2$Kavli IPMU (WPI), UTIAS, the University of Tokyo, Kashiwa, Chiba 277-8583, Japan\\
}

\date{\today}

\begin{abstract}
We consider on one hand the possibility that  a supersymmetric ${\cal N}=1$ conformal gauge theory has a strongly coupled locus on the conformal manifold at which a different, dual, conformal gauge theory becomes a good weakly coupled description. On the other hand we discuss the possibility that strongly coupled  theories, {\it e.g.} SCFTs in class ${\cal S}$, having exactly marginal  ${\cal N}=1$ deformations admit a weakly coupled gauge theory description on some locus of the conformal manifold. We present a simple algorithm to search for such dualities and discuss several concrete examples. In particular we find conformal duals for  ${\cal N}=1$  SQCD models with $G_2$ gauge group and a model with $SU(4)$ gauge group in terms of simple quiver gauge theories. We also find conformal weakly coupled quiver theory duals for a variety of class ${\cal S}$ theories: $T_4$, $R_{0,4}$, $R_{2,5}$, and rank $2n$ Minahan-Nemeschansky $E_6$ theories. Finally we derive conformal Lagrangians for four dimensional theories obtained by compactifying the E-string on genus $g>1$ surface with zero flux.
The pairs of dual Lagrangians  at the weakly coupled loci have different symmetries which are broken on a general point of the conformal manifold. We match the dimensions of the conformal manifolds, symmetries on the generic locus of the conformal manifold, anomalies, and supersymmetric indices.   The simplicity of the procedure suggests that such dualities are ubiquitous.

\end{abstract}


\maketitle

\section{Introduction}

Let us consider conformal ${\cal N}=1$ supersymmetric gauge theories in four dimensions. Here by conformal ${\cal N}=1$ gauge theories we shall mean gauge theories with a conformal manifold passing through zero gauge couplings. A variety of such models have been widely studied with the important examples being ${\cal N}=4$ SYM and, say, ${\cal N}=2$ $SU(N)$ $N_f=2N$ SQCD.  To build such theories one should first choose the matter content so that the one loop beta functions for all the gauge couplings and the gauge anomalies will vanish. Turning on the gauge coupling by itself is a marginally irrelevant deformation. However, when supplemented with superpotential terms in certain cases one can construct exactly marginal deformations parametrizing the conformal manifold ${\cal M}_c$ of the theory. The dimension of the conformal manifold, $dim {\cal M}_c$ can be computed using a variety of techniques \cite{Leigh:1995ep,Green:2010da}.

Theories residing at different points of the conformal manifold are different SCFTs. For example generic correlation functions computed for such models vary with the position on the conformal manifold. However some quantities are invariants of such a position. These quantities involve conformal anomalies $a$ and $c$, supersymmetric protected quantities (which often can be encoded in different indices), symmetry on generic points of the conformal manifold $G_F$, and 't Hooft anomalies  for these symmetries. The symmetry of the theories can enhance on sub-loci of the manifold, however the symmetry on a generic point is expected to be invariant. We will refer to the properties of SCFTs which do not change on the conformal manifold as the {\it ${\cal M}_c$ invariants}.

An interesting question about such conformal maniflolds is whether cranking up the coupling constants one can arrive at special loci  where an alternative weakly coupled description is more suitable. This phenomenon is known as conformal duality. Well known examples include the ${\cal N}=4$ SYM and ${\cal N}=2$ $SU(N)$ $N_f=2N$ SQCD.

The dual theory can be again a gauge theory or it can be a more abstractly defined strongly coupled SCFT \cite{Argyres:2007cn,Gaiotto:2009we}. 
Alternatively, we know of a huge variety of strongly coupled SCFTs which upon conformally gauging some symmetries are dual to usual gauge theories. A natural question is then whether before gauging the global symmetries these SCFTs reside on a conformal manifold of some conformal gauge theory. A necessary condition for this is that these SCFTs admit exactly marginal deformations. 

In this paper we give a plethora of novel examples of dualities of this type. We discuss both non trivial conformal dualities between gauge theories, and dualities between known strongly interacting SCFTs and simple gauge theories. In fact we present a very simple algorithm to search for such dualities. We stress that a given model of the type discussed here might or might not have a simple conformal  dual, however applying the algorithm we find that the conformal dualities are quite ubiquitous.

The structure of this article is as follows. We start with a brief description of the algorithm in section two. In  section three we apply it to construct conformal duals of simple gauge theories. In section four we discuss strongly coupled ${\cal N}=2$ SCFTs, and in section five we discuss ${\cal N}=1$ SCFTs obtained by compactifications of the rank one E-string theory. We conclude in section six with some general comments.

\section{The basic idea}

Given a conformal theory $T_1$ with conformal manifold ${\cal M}_c$ and the ${\cal M}_c$ invariants we can systematically seek for a dual conformal gauge theory $T_2$ which might reside on the same conformal manifold. Such a theory might or might not exist, however if it does exist the properties of this model are severely constrained. First, we look at conformal anomalies of $T_1$, which is part of the invariant information, and define,
\be
&&a= a_v n_v+a_\chi n_\chi\,,\quad c= c_v n_v+c_\chi n_\chi\,.
\ee  Here we define the contribution to the conformal anomalies of vector and chiral fields as $(a_v,c_v)=(\frac3{16},\frac18)$ and $(a_\chi,c_\chi)=(\frac1{48},\frac1{24})$. The numbers $n_v$ and $n_\chi$ are the effective numbers  of vectors and chirals which the theory $T_1$ has. The dual conformal gauge theory should have these numbers of vector and chiral fields. If $a$ and $c$ determine these numbers to be not integer a conformal dual gauge theory cannot exist. Also $n_v$ must be the sum of dimensions of non-abelian gauge groups, which is quite restrictive for small $n_v$. Next, we search over all the conformal gauge theories with the given $n_v$ vectors and $n_\chi$ chiral fields for models such that all the gauge couplings have vanishing one loop $\beta$ functions. The number of possibilities to search through  is finite and thus we will find some finite number of models which satisfy this constraint. Next, we should verify that these models have a conformal manifold and that its dimension, $dim {\cal M}_c$, and symmetry on a generic locus, $G_F$, match the ones of $T_1$. The computation of the dimension and the symmetries is most efficiently done by listing all the marginal operators $\lambda_\alpha$ at the free point, determining the symmetry at the free point $G_{free}$, and then computing the Kahler quotient $\{\lambda_\alpha\}/G_{free}^{\mathbb C}$ \cite{Green:2010da} (see also \cite{Kol:2002zt}). Note that in principle we need to include in this counting gauge couplings and anomalous symmetries, but these typically cancel each other in the quotient.
Finally, we should match all the remaining invariant information which includes at the minimum 't Hooft anomalies for $G_F$ and the protected spectrum. If a model satisfying all these is found it is a candidate for a dual description.  We stress that this very systematic algorithm is not guaranteed to produce a dual theory, as such a conformal gauge theory might simply not exist, but if it does exist the algorithm will find it. Surprisingly we do find that many theories have such a conformal gauge theory dual.

We will apply this algorithm in three cases. First, we will search through low values of $n_v$ for conformal gauge theories with a simple gauge group. Second, we will consider some strongly interacting ${\cal N}=2$ theories of class ${\cal S}$ \cite{Gaiotto:2009we,Gaiotto:2009hg} obtained by compactifications of $A_{N-1}$ $(2,0)$ theories on Riemann surfaces with low values of $N$ and seek for dual conformal gauge theories for these. Third, we consider strongly interacting ${\cal N}=1$ theories obtained by compactifications of the $6d$ rank $1$ E-string SCFT on Riemann surfaces. Before discussing the examples let us make some general comments.

Here we consider various examples of dualities between two conformal theories that arise at different points on an $\mathcal{N}=1$ conformal manifold. The examples fall into three types. In the first type both dual theories are Lagrangian $\mathcal{N}=1$ conformal gauge theories. We want to consider the simplest case so we take one side to be just one simple gauge group with matter such that the one loop beta function vanishes and there is a conformal manifold. Ideally we would like the other side to also be a single simple gauge group, however, in general it is hard to achieve this. This follows as the two dual theories must have the same $a$ and $c$ conformal anomalies, which for conformal gauge theories translates to equal numbers of vectors and chiral fields. Generally the dimension of each group is unique with the general exception of $USp(2N)$ and $SO(2N+1)$ and a few sporadic ones like $E_6$, $USp(12)$ and $SO(13)$. As a result we do not usually expect a simple gauge group to be dual to another simple group unless it is the exact same group, modulo a few exceptions. The case of $\mathcal{N}=4$ SYM with gauge groups $USp(2N)/SO(2N+1)$ is an example realizing one of these exceptions.

The second type of duality we consider has on one side a Lagrangian $\mathcal{N}=1$ gauge theory while on the other side we have an $\mathcal{N}=2$ SCFT. Here we shall concentrate on the more interesting case where the $\mathcal{N}=2$ SCFT does not have a manifestly $\mathcal{N}=2$ Lagrangian, though in principle the same type of dualities may also be found in Lagrangian $\mathcal{N}=2$ SCFTs. First, for such a duality to be possible the $\mathcal{N}=2$ SCFT must have an $\mathcal{N}=1$ only preserving conformal manifold. Such conformal manifolds for $\mathcal{N}=2$ SCFTs are relatively unstudied so we first would like to address under what conditions do these appear. First, consider the $\mathcal{N}=2$ SCFT as an $\mathcal{N}=1$ SCFT. In that view point, the SCFT has an $U(1)_t \times G$ global symmetry where $G$ is the $\mathcal{N}=2$ flavor symmetry and $U(1)_t$ is the commutant of the $\mathcal{N}=1$ $U(1)_{\hat R}$ R-symmetry in the $\mathcal{N}=2$ $U(1)_r \times SU(2)_R$ R-symmetry. In Lagrangian theories, it is the symmetry that acts on the adjoint chiral in the $\mathcal{N}=2$ vector multiplet, let us denote it as $\Phi$, with charge $-1$ and on the chiral fields in the hypermultiplet, let us denote them as $\chi_1, \chi_2$, with charge $\frac{1}{2}$. In order to have an $\mathcal{N}=1$ only preserving conformal manifold we must have marginal operators with a non-trivial Kahler quotient under both $U(1)_t$ and $G$.

We next want to examine what types of marginal operators we can expect for $\mathcal{N}=2$ SCFTs. For simplicity we consider here only Lagrangian theories. We expect this to also hold for non-Lagrangian theories that can be related to Lagrangian theories via gauging some of their symmetries, which will be the ones we consider in this article. It might be interesting to study this using $\mathcal{N}=2$ superconformal representation theory, but we reserve this for future study. The marginal operators then are those built from three chiral fields. We can generally separate them to three classes: Coulomb branch operators, Higgs branch operators, and mixed branch operators. The first class, Coulomb branch operators, are the ones built solely from the adjoint chiral in the $\mathcal{N}=2$ vector multiplet, and are usually of the form $Tr(\Phi^3)$ in Lagrangian theories. More generically, these are dimension three Coulomb branch operators. They have charge $-3$ under $U(1)_t$ and are uncharged under $G$.

The second class are the Higgs branch operators, which in Lagrangian theories are dimension three operators built solely from chiral fields in the hypermultiplets. In more general theories they are Higgs branch chiral ring operators of dimension three. These have charge $\frac{3}{2}$ under $U(1)_t$ and are usually charged under $G$ in some self-conjugate representation. The last class is that of mixed branch operators. In Lagrangian theories these are dimension three operators built from chiral fields in both the vector and hyper multiplets. Probably the most well known of these are the $\mathcal{N}=2$ preserving marginal operators that exist in conformal $\mathcal{N}=2$ gauge theories, which are of the form $\chi_1 \Phi \chi_2$. Besides these, there can be additional operators of this type either in the form of $\chi \Phi \chi$ or $\Phi^2 \chi$. As an example of an SCFT with the former we have the conformal $\mathcal{N}=2$ $SU(N)$ gauge theory with one hypermultiplet in the antisymmetric representation and one hypermultiplet in the symmetric representation, which has marginal operators of the form $\chi_S \Phi \chi_{\overline{AS}} + \chi_{AS} \Phi \chi_{\overline{S}}$. As an example of the latter we have any conformal $\mathcal{N}=2$ $USp$ type gauge theory with hypermultiplets in the antisymmetric representation, which has marginal operators of the form $\Phi^2 \chi_{AS}$. These type of operators have charge $0$ and $-\frac{3}{2}$, respectively, under $U(1)_t$ and are charged under $G$ in some self-conjugate representation.

As previously mentioned to have the $\mathcal{N}=1$ only preserving conformal manifold we need the marginal operators to have a non-trivial Kahler quotient with respect to $U(1)_t$ and $G$. Since the representations under $G$ are self-conjugate it is not unreasonable for there to be a Kahler quotient under it. However, the $U(1)_t$ charges are not coming in pairs of opposite charges, and getting a Kahler quotient under it is quite non-trivial. This requires having either the mixed branch operators with charge $0$ or Higgs branch operators and either Coulomb branch or the charged mixed branch operators. Here we shall concentrate on cases having both Higgs branch and Coulomb branch marginal operators. The reason for this is that by now there are known techniques to extract these for the class of non-Lagrangian theories in class ${\cal S}$. Specifically, there are known methods to extract the dimension of Coulomb branch operators for class ${\cal S}$ theories, see for instance \cite{CD}, and the Higgs branch operators can be extracted from the Hall-Littlewood index for which there are known expressions for class S theories, see \cite{GR}. Alternatively, to our knowledge, there is no systematic way to extract mixed branch operators save for trying to infer them from dualities with Lagrangian theories.

Finally, in the third type of duality, one side is a Lagrangian $\mathcal{N}=1$ gauge theory while on the other side we have an $\mathcal{N}=1$ strongly coupled SCFT.
Similarly to the $\mathcal{N}=2$ case, we can use compactifications of $6d$ SCFTs on Riemann surfaces to generate interesting examples of such theories. The specific case of a genus $g>1$ Riemann surface without punctures or flux is especially appealing for several reasons. First, these class of theories are expected to have a large conformal manifold, related to complex structure deformations of the Riemann surfaces and flavor holonomies, on a generic point of which the global symmetry is completely broken \cite{Benini:2009mz,Razamat:2016dpl}. Also, the $6d$ construction allows us to compute various quantities of interest like the $a$ and $c$ central charges, see for instance \cite{Benini:2009mz,Razamat:2016dpl}, and for this class of compactifications these are guaranteed to be rational. As a result, it is possible that some of these have dual $\mathcal{N}=1$ conformal gauge theories. 

\

\section{Conformal  duals of ${\cal N}=1$ gauge theories}

Let us start by considering the duals of conformal gauge theories with a simple gauge group.
We will only consider models with minimal ${\cal N}=1$ supersymmetry although the search can be done also for theories with extended supersymmetry.  To do so one can systematically scan through various values of $n_v$. The smallest possible value  is $n_v=3$, however we do not have conformal ${\cal N}=1$ $SU(2)$ gauge theories. Next, we can consider $n_v=8$ and here one already can find several examples, see \cite{Classification} for a classification. For example $SU(3)$  SQCD with nine flavors,  $SU(3)$ with chiral fields in ${\bf 6}\oplus {\bf \overline 6}\oplus (4\times {\bf 3})\oplus (4\times {\bf \overline 3})$ are conformal. However all these models have different $n_\chi$ and so cannot be dual to each other. Moreover as $8$ is not divisible by $3$ we cannot construct a quiver theory with $SU(2)$ gauge groups to have the same $n_v$. Thus we conclude that $SU(3)$ gauge theories, if they have a conformal gauge theory dual on the conformal manifold, have to be self-dual. Increasing $n_v$ the next value is $10$ for the group $USp(4)$ and here as for $n_v=8$ the only possibility is self duality. The next case is $G_2$, with $n_v=14$. Here we have two conformal ${\cal N}=1$ gauge theories, one with matter in $3\times {\bf 7}\oplus {\bf 27}$ and another with matter in $12\times {\bf 7}$. As $14=3+3+8$, we can have duals with $SU(2)^2\times SU(3)$ gauge groups and we will proceed to discuss these cases in detail. The next possible value of $n_v$ is $15$ for $SU(4)$. We have a large variety of conformal gauge theories with $SU(4)$ gauge group. As $15=3\times 5$ these might have duals with $5$ $SU(2)$ gauge groups and we will discuss one such example.

\subsection{Dual of ${\cal N}=1$ $G_2$ SCFT with $3\times {\bf 7}\oplus {\bf 27}$}

Let us consider ${\cal N}=1$ SQCD with gauge group $G_2$, three fundamentals $Q_i$, and one chiral field in the ${\bf 27}$, $\widetilde Q$. The one loop beta function of this model vanishes implying that the superconformal R charges of all the chiral fields are $\frac23$. The symmetry at the free point is $U(1)\times SU(3)$. The fundamentals $Q_i$ are a triplet of $SU(3)$ and have $U(1)$ charge $-3$ and the ${\bf 27}$ charge $+1$. 
 This theory has a number of marginal operators.  Note that the ${\bf 7}$ has an antisymmetric cubic invariant using which we can build a marginal superpotential. Since we have three fundamental fields this gives rise to one such term, $\epsilon^{ijk}Q_iQ_jQ_k$, which is a singlet of $SU(3)$ and has $U(1)$ charge $-9$. The ${\bf 27}$ has two independent symmetric cubic invariants giving rise to two marginal operators, we denote them as $(\widetilde Q)^3_1$ and  $(\widetilde Q)^3_2$, which are singlets of $SU(3)$ and have $U(1)$ charge $+3$.  Finally we can build marginal operators as $\widetilde Q Q_{(i} Q_{j)}$, which have $U(1)$ charge $-5$ and are in the ${\bf 6}$ of $SU(3)$. We can now compute the dimension of the conformal manifold \cite{Green:2010da} by computing the Kahler quotient generated by the marginal couplings. Just considering $\epsilon^{ijk}Q_iQ_jQ_k$, $(\widetilde Q)^3_1$, $(\widetilde Q)^3_2$, and the gauge coupling we can build two independent singlets giving rise to two exactly marginal directions. We can build one independent singlet of $SU(3)$ from the marginal coupling of $\widetilde Q Q_{(i} Q_{j)}$ which gives rise to an additional exactly marginal coupling. We thus deduce that the theory is conformal and has a three dimensional manifold of exactly marginal couplings.  Note that the $U(1)$ symmetry is broken by all the exactly marginal deformations while $SU(3)$ is preserved by the first two. The last deformation breaks $SU(3)$ down to $SO(3)$ which is the symmetry preserved at a generic point of the conformal manifold.
 
\begin{figure}[htbp]
\includegraphics[scale=0.62]{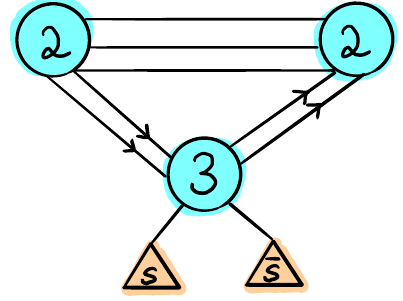}
\caption{The dual of $G_2$ with $3\times {\bf 7}\oplus {\bf 27}$.  In this and the following figures one should think of the models with all the possible gauge invariant cubic superpotentials turned on. }\label{F:G23}
\end{figure}

Finally let us mention that the conformal anomalies of this model are,
\be
a=14 \, a_v+48 a_\chi =\frac{29 }8\, , \quad c= 14 \, c_v+48 c_\chi =\frac{15 }4\,.
\ee 
Let us seek a conformal dual of this theory. We are after a theory with $14$ vectors and $48$ chirals. As previously mentioned, besides $G_2$ we can have $14$ vector multiplets also from two $SU(2)$ gauge groups and one $SU(3)$ group, and this is the only possibility. Now we need to make sure the one loop beta function of each gauge group vanishes and that the total number of chiral fields is $48$. One can accomplish this and the result is depicted in Figure \ref{F:G23}. Here $s$ and $\overline s$ stand for the ${\bf 6}$ and ${\bf \overline 6}$ representations of $SU(3)$. By construction the conformal anomalies of this model agree with the $G_2$ SQCD. We mention that this choice of matter content is not the only one which satisfies matching the anomalies and vanishing beta functions, for example orienting differently some of the arrows will do this but will give an inequivalent model. However, we claim as will be discussed below, that the quiver in  Figure \ref{F:G23} is the dual to the $G_2$ SQCD.
Let us now analyze the symmetries and  the conformal manifold. The theory at the free point has symmetry $SU(3)\times SU(2)^2\times U(1)^2$. The $U(1)^2$ charges of  the bifundamentals between the two $SU(2)$s are $(+1,0)$,  between the $SU(2)$s and the $SU(3)$ $(-1,0)$, and of $s$ and  $\overline s$ are $(\frac45,\pm1)$ respectively.

\begin{figure}[htbp]
\includegraphics[scale=0.62]{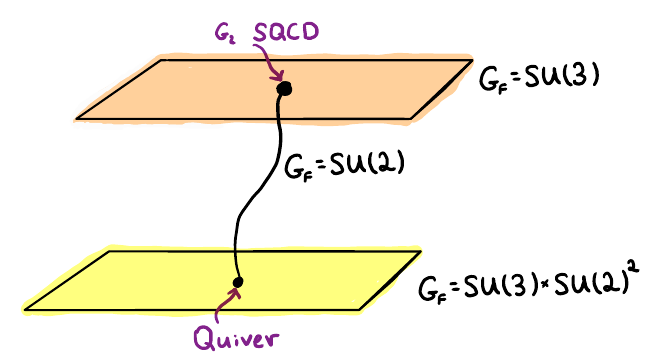}
\caption{The structure of the conformal manifold.}\label{F:Confman}
\end{figure}

 There are three types of  marginal operators. The First one corresponds to the triangles in the quiver and these are in the representation $({\bf 3},{\bf 2},{\bf 2})_{(-1,0)}$. The second one correspond to  the cubic invariants of the symmetric and conjugate symmetric
 which have charges $(1,1,1)_{( \frac{12}{5},\pm 3)}$. The third one are the (conjugate) symmetric times the square of the bifundamentals between $SU(3)$ and the two $SU(2)$s with charges $(1,1,1)_{(-\frac65,\pm1)}$. The last two types of deformations have a non trivial Kahler quotient of dimension two along which the two $U(1)$ symmetries are broken but $SU(3)\times SU(2)^2 $ is preserved. Then we can build an additional exactly marginal operator using the first type of deformations. This will break all the symmetry but the diagonal combination of the two $SU(2)$s and $SO(3)$ in $SU(3)$. All in all, as above we get a three dimensional conformal manifold with $SU(2)$ symmetry preserved on a general locus. We have a two dimensional locus with enhanced $SU(3)\times SU(2)^2$ symmetry. Note that both duality frames have two dimensional loci with enhanced symmetry which are however different.\footnote{To be more precise we solve here the problem of K\"{a}hler quotient by using only the non-anomalous symmetries and the solution of this problem has the structure shown in Figure \ref{F:Confman}. However, the inclusion of the anomalous symmetries in this particular case is subtle and will imply that the solution of the K\"{a}hler quotient of the quiver theory preserving the non-Abelian symmetry is only one dimensional and  has some of the gauge couplings squared negative along the second naive direction. This issue does not play a role in the duality we discuss and thus we will be cavalier about it. We thank Z. Komargodski for discussions of such issues.}
  This is not a contradiction of the duality as the two do not have to intersect. See Figure \ref{F:Confman}.
 
The only  anomalies we need to match are the ones for symmetries on general points of the conformal manifold. As we already matched the conformal anomalies, which implies matching of R-symmetry anomalies, the only anomaly left is $Tr U(1)_R SU(2)^2$. In the $G_2$ model the $SO(3)$ is imbedded in $SU(3)$ and we have one triplet in the fundamental of $G_2$. This gives us then $Tr U(1)_R SU(2)^2=(-\frac13)\times 7\times 2$. On the quiver side the $SU(2)$ is the diagonal of the two $SU(2)$s in the quiver and $SO(3)\in SU(3)$. This gives us 

\be
TrU(1)_R SU(2)^2=(-\frac13)(6\times \frac12+6\times \frac12+4\times 2)\,.
\ee 
We see that this anomaly precisely matches.

Finally we can match the supersymmetric indices \cite{Kinney:2005ej}.  The index computed at the free point of the $G_2$ theory is,
 \be
&&I  =  1 +(p q)^{\frac{2}{3}}(x^2 + \frac{1}{x^6} \bold{6}_{SU(3)}) \\
&& + p q\biggl(2x^3 + \frac{1}{x^5} \bold{6}_{SU(3)} + \frac{1}{x^9} - \bold{8}_{SU(3)} - 1\biggr) +\cdots \,.\nonumber\ee 
Here we use the standard notations \cite{index} for the index and $x$ is the fugacity for the $U(1)$.  The index at the free point of the quiver theory is,
\be
&&I  =  1 +(p q)^{\frac{2}{3}}(a^{\frac85} + a^{2} \bold{6}_{SU(3)})+ p q\biggl(a^{\frac{12}5}(b^3 + \frac{1}{b^3})+\nonumber \\
&&  \frac1a \bold{2}_{SU(2)_1}\bold{2}_{SU(2)_2}\bold{3}_{SU(3)} + \frac1{a^{\frac65} }(b + \frac{1}{b}) \nonumber -  \bold{3}_{SU(2)_1} -\nonumber\\
&& \bold{3}_{SU(2)_2} - \bold{8}_{SU(3)} - 2\biggr)  + \cdots. \label{IndexThrBDual1}
\ee  Here $a$ and $b$ are the fugacities for the two $U(1)$s. Note that specializing to symmetries preserved on a generic locus of the conformal manifold, that is $a=b=x=1$ and ${\bf 3}_{SU(3)}=\overline {\bf 3}_{SU(3)}={\bf 3}_{SO(3)}$ for $G_2$ and ${\bf 3}_{SU(2)_1}={\bf 3}_{SU(2)_2}={\bf 3}_{SU(3)}={\bf 3}_{SO(3)}$ for the quiver, the two indices precisely agree,
\be
I =1+ (p q)^{\frac{2}{3}}(2 + \bold{5}) + p q(3 - \bold{3})+ \cdots . \label{IndexThrADual1}
\ee This can be checked to rather high order in the expansion in terms of the fugacities. We thus have compelling evidence that in fact the two models are conformally dual to each other. That is that there should be a map between the conformal manifolds of the two models which will describe equivalent theories. The manifold has at least two cusps at which one of the two models is weakly coupled.

An additional simple check of the duality we can perform is to study RG flows. We can only compare flows which exist on a generic point of the conformal manifold. One such flow is giving a vacuum expectation value to one of the ${\bf 7}$ on the $G_2$ side. On the dual side this corresponds to giving a vev to one of the three bifundamentals between the two $SU(2)$s. Let us first analyze the flow in the latter frame. Giving the vev locks the two $SU(2)$ gauge groups together, and gives a mass to the remaining bifundamentals between the two $SU(2)$s and to two out of the four bifundamentals between the $SU(2)$s and the $SU(3)$. The remaining two acquire R-charge $\frac13$. The theory in the IR is just an $SU(2)\times SU(3)$ gauge theory with two bifundamentals and the symmetric and conjugate symmetric for the $SU(3)$. The $SU(2)$ gauge node has three flavors and thus flows in the IR to a Wess-Zumino model with $15$ gauge singlet fields \cite{Seiberg:1994pq}. The R-charge of these fields is $\frac23$ and in terms of $SU(3)$ the representations are $1+{\bf 8}+{\bf 3}+{\bf \overline 3}$. Thus in the end we get a conformal theory which consists of a single decouple chiral field and ${\cal N}=2$ $SU(3)$ gauge theory with one fundamental and one symmetric hypermultiplet. On the $G_2$ side giving a vev to one of the ${\bf 7}$s breaks the gauge group to $SU(3)$. The remaining two ${\bf 7}$s get a mass and the ${\bf 27}$ decomposes under the remaining gauge $SU(3)$ as ${\bf 27}=1+{\bf 3}+{\bf \overline 3}+{\bf 6}+{\bf \overline 6}+{\bf 8}$. Thus in the end we get manifestly the same model as in the dual frame. This is yet another direct check of the proposed duality.

We can also consider mass deformations to get IR dualities, in the sense of \cite{Seiberg:1994pq}, from this conformal duality though we shall not analyze this in detail here.

\

\subsection{Dual of ${\cal N}=1$ $G_2$ SCFT with $12\times {\bf 7}$}

Let us now consider ${\cal N}=1$ SQCD with gauge group $G_2$ and $12$ fundamentals $Q_i$.\footnote{The IR dualities of $G_2$ with $N_f<12$ fundamentals were discussed by Pouliot in \cite{Pouliot:1995zc}.}  This theory is conformal. The symmetry group at the free point is $SU(12)$. The marginal operators are built from the antisymmetric cubic invariant of the fundamentals. One can perform the Kahler quotient and find that on the conformal manifold all the symmetry is broken and thus the number of exactly marginal operators is the number of marginal operators minus the number of currents, $\frac16 10\time 11\times 12-12^2+1=77$.

 The conformal anomalies are,
\be
a=14 \, a_v+84 a_\chi =\frac{35 }8\, , \quad c= 14 \, c_v+84 c_\chi =\frac{21 }4\,.
\ee The relevant operators are built from the symmetric square of the fundamentals and their number is $78$.

\begin{figure}[htbp]
\includegraphics[scale=0.62]{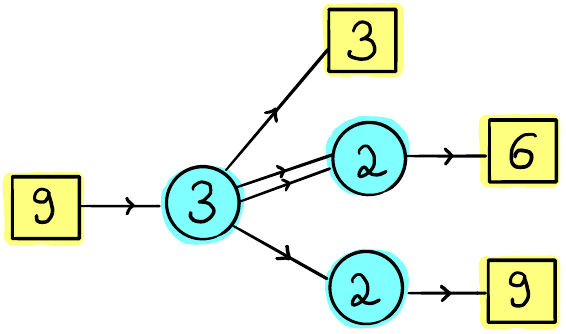}
\caption{The dual of $G_2$ with $12\times {\bf 7}$. }\label{F:G212}
\end{figure}

We now look for a conformal dual gauge theory. As in the previous example the only possible gauge group is $SU(3)\times SU(2)^2$. An example of a quiver with the correct number of chiral fields and vanishing one loop beta function is in Figure \ref{F:G212}. We can next count the number of relevant operators. We have $9\times 3$ from mesons of the $SU(3)$ gauge group, $\frac12 6\times 5$ from flavor gauge invariants of the upper $SU(2)$ and $\frac12 9\times 8$ from flavor gauge invariants of the lower $SU(2)$. All in all, we get $3\times 9+\frac126\times 5+\frac12 9\times 8=78$ matching the $G_2$ side.

Next we count the marginal operators. First we have the gauge invariants built from  operators winding the quiver, whose number is $9\times 9+9\times 2\times 6$. Second we have baryonic operators of the $SU(3)$ gauge group which give rise to $84+12+1$ marginal operators. All in all we get $286$ marginal operators. The non-anomalous global symmetry at the free point is $SU(9)^2\times SU(6)\times SU(3)\times SU(2)\times U(1)^3$ which gives $209$ conserved currents. Computing the Kahler quotient here also all the symmetry is broken on a generic locus of the conformal manifold giving rise to $77$ exactly marginal deformations, which agrees perfectly with the $G_2$ SQCD.

We can compare the indices switching off all the fugacities for global symmetries as these are absent on the general point of the conformal manifold, and obtain in both cases the same result,
\be
&&{\cal I} = 1+78 (q p)^{\frac23} +77 qp +78 (q p)^{\frac23} (q+p)+2850 (qp)^{\frac43}+\nonumber\\
&&76 qp (q+p)+4446(q p)^{\frac53}+78(q p)^{\frac23} (q^2+p^2)+\cdots\,.
\ee
We can  check the matching to relatively high order in expansion in fugacities. 

\

\subsection{Dual of ${\cal N}=1$ $SU(4)$ SCFT with $4\times {\bf 6}\oplus8\times {\bf 4}\oplus 8\times {\bf \overline 4}$ }

So far we have considered conformal duals for gauge group $G_2$, but here we would like to try and consider at least one other group. One reason for this is a desire to explore how common are these types of dualities. For this it is useful to try at least one more case. To try to keep things simple, we shall consider the group $SU(4)$ whose dimension, $15$, comes right after $G_2$. Like $G_2$ there is only one other combination of simple groups with the same number, five $SU(2)$ groups, and we shall  seek a dual with this vector content.

Next we need to choose a conformal $SU(4)$ gauge theory. First, we can consider the case with just fundamentals, specifically, $12$ fundamental and $12$ antifundamental chirals. While with this combination the one loop beta function at the free point vanishes, there are no marginal operators one can turn on so this theory is actually IR free. Next then, we can consider the case with fundamentals and antisymmetric representations. Here we do have marginal operators connecting the antisymmetric chirals to a pair of fundamental or antifundamental chirals. However, we can consider the  $U(1)$ acting on the antisymmetrics with one sign and on the fundamentals and antifundamentals with another with charges chosen so that it is non-anomalous. The marginal operators are all going to have the same charge under this symmetry and so there is no Kahler quotient, unless the marginal operators happen to be uncharged under it. There is precisely one choice that is consistent with gauge anomaly cancellation, vanishing of the one loop beta function and existence of a Kahler quotient. The miraculous combination is four antisymmetrics, eight fundamentals and eight antifundamentals.  

\begin{figure}[htbp]
\includegraphics[scale=0.62]{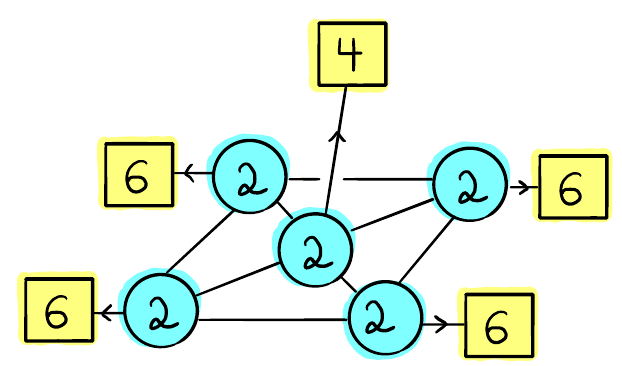}
\caption{The dual of $SU(4)$ with $4\times {\bf 6}\oplus8\times {\bf 4}\oplus 8\times {\bf \overline 4}$. }\label{F:Su4}
\end{figure}

This leads us to consider an $SU(4)$ gauge theory with four chiral fields in the ${\bf 6}$ and eight fundamental flavors $({\bf 4},{\bf \overline 4})$. One can perform the Kahler quotient and conclude that this theory is indeed conformal with an $82$ dimensional conformal manifold. On a generic point on the conformal manifold all the symmetry is broken, except the previously mentioned $U(1)$, under which the marginal operators are uncharged enabling the quotient. Under this $U(1)$ symmetry the fundamentals and antifundamentals have charge $+1$ and the ${\bf 6}$s are charged $-2$. The model has $10$ relevant operators built from ${\bf 6}$s with charge $-4$ and 64 mesons with charge $+2$. The conformal anomalies of the model are
\be
a=15 \, a_v+88 a_\chi =\frac{223 }{48}\, , \quad c= 15 \, c_v+88 c_\chi =\frac{133 }{24}\,.
\ee   
As previously mentioned, we seek a dual for this theory with five $SU(2)$ gauge groups. The quiver is in Figure \ref{F:Su4}. Let us define here a $U(1)$ 
symmetry which assigns charge $+1$ to the bifundamentals of $SU(6)$s and $SU(2)$s and to bifundamentals of the perimeter $SU(2)$s and the center $SU(2)$, and charge $-2$ to the rest of the fields. Note that then all the marginal operators have zero charge. The conformal manifold of this model is again $82$ dimensional and has a preserved $U(1)$. The index of the two models perfectly agrees in an expansion in fugacities with the refinement for the preserved $U(1)$ with fugacity $b$,
\be
&&{\cal I}=1+(qp)^{\frac23}(1+q+p)\left(10 b^{-4}+64 b^2\right)+81 qp +\\
&&\left(55b^{-8}+576 b^{-2}+2002 b^4\right) (qp)^{\frac43}+80(q+p)qp+\cdots\,. \nonumber
\ee
 We should also verify the anomalies for the $U(1)$. 
In both theories we have $64$ chiral fields with free  R charge and $U(1)$ charge $+1$ and $24$ chiral fields with free R charge and $U(1)$ charge $-2$. The anomalies thus manifestly match.

\section{Conformal duals of class ${\cal S}$ theories}

Let us now apply the algorithm to search for dual conformal gauge theories for some of the strongly coupled SCFTs in class ${\cal S}$ of type $A_{N-1}$. Many of the SCFTs we discuss here and their properties appear in \cite{CD}. We will concentrate on the simplest models corresponding to spheres with three punctures. We can organize the search by increasing the value of $N$. For $N=2$ the theories are quivers with $SU(2)$ gauge groups and we do not have any strongly coupled SCFTs. For $N=3$ the model $T_3$ is strongly coupled but it does not have a conformal manifold and thus our technology does not apply. For $N=4$ however we have several interesting models. The $E_7$ MN theory again does not have a conformal manifold and thus we cannot discuss it, but the $R_{0,4}$ and the $T_4$ models do and we will consider them. We will also discuss one example for $N=5$, the $R_{2,5}$ model, and a sequence of models with $N=6n$, the so call rank $2n$ $E_6$ MN theories.

All the examples we discuss have both dimension three Coulomb and Higgs branch operators suggesting the existence of a conformal manifold. A priori they may also have mixed branch operators. Besides the $R_{0,4}$ model, where the full index is known \cite{Agarwal:2018ejn}, we do not know how to systematically extract these for the other models. In some cases, like the $R_{2,5}$ model, one can use dualities with gauge theories to infer these. Here, when matching the conformal manifold, we shall assume these are not present.  

\subsection{Dual of ${\cal N}=2$ $R_{0,4}$ SCFT}

Let us now consider the $R_{0,4}$ ${\cal N}=2$ SCFT. This is a strongly coupled model which can be obtained in class ${\cal S}$ construction as a compactification of $A_3$ $(2,0)$ theory on a sphere with two maximal and one next to maximal puncture \cite{CD}. In particular, turning on certain vevs it flows to the Minahan-Nemeschansky $E_7$ SCFT.  Several facts are known about this model. In particular the conformal anomalies are,
\be
a=12 \, a_v+72 a_\chi =\frac{15 }4\,, \quad c= 12 \, c_v+72 c_\chi =\frac{9 }2\,.
\ee The model has $SU(8)\times SU(2)$ global symmetry.
The spectrum of protected operators is also known. For example it was computed in \cite{Agarwal:2018ejn} using a construction with a singular Lagrangian. In such constructions one starts with a Lagrangian and then gauges a symmetry which only appears at a strongly coupled cusp, see  \cite{Gadde:2015xta}. The result is,
\be 
&&{\cal I}=1+t({\bf 63}_{SU(8)}+{\bf 3}_{SU(2)}) (pq)^{\frac23}+\biggl(t^{\frac32} {\bf 70}_{SU(8)}{\bf 2}_{SU(2)}\nonumber\\
&&+\frac1{t^3}-1-{\bf 63}_{SU(8)}-{\bf 3}_{SU(2)}\biggr) q p+\cdots\,.
\ee In ${\cal N}=1$ language the theory has an additional $U(1)_t$ symmetry coming from the enhanced R-symmetry. From the index we can read off the marginal operators as being in
representation ${\bf 70}_{SU(8)}{\bf 2}_{SU(2)}$ with $U(1)_t$ charge $\frac32$ and a singlet with charge $-3$. The latter is the dimension three Coulomb branch operator that is known to exist in this theory while the former is a dimension three Higgs branch operator. Using these operators we can construct $74$ exactly marginal directions with all the symmetry broken along the general direction.  All the marginal deformations break ${\cal N}=2$ to ${\cal N}=1$.

\begin{figure}[htbp]
\includegraphics[scale=0.62]{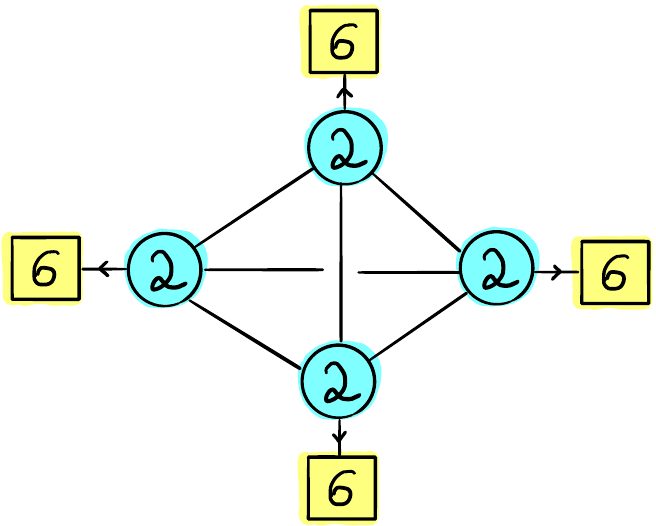}
\caption{The ${\cal N}=1$ conformal dual of $R_{0,4}$.}\label{F:R04}
\end{figure} 
To construct a conformal dual we are after a theory with $12$ vectors and $72$ chiral fields. The unique possibility with $12$ vectors is four $SU(2)$ groups. The quiver in Figure \ref{F:R04} has the right matter content and all gauge groups have vanishing one loop beta function. The model at the free point has $SU(6)^4\times U(1)^{6}$ non-anomalous global symmetry. The marginal deformations correspond to triangles for each of the $4$ faces of the quiver, and to cubic gauge singlets starting and ending at two different $SU(6)$ global symmetry groups. All in all we have $220$ marginal operators and $146$ conserved currents. Computing the Kahler quotient we get precisely $74$ exactly marginal deformations and on a generic point of the conformal manifold all the symmetry is broken. The index of this theory at a generic point, that is switching off fugacities for global symmetries, is
\be
{\cal I}=1+66 (qp)^{\frac23}+74 qp+\cdots\,,
\ee which precisely agrees with the result for $R_{0,4}$. This can be easily verified to higher orders in expansion in fugacities. We thus can conjecture that the tetrahedral quiver and $R_{0,4}$ are dual to each other. That is they are descriptions at two different points on the same conformal manifold.

It is instructive the study the Kahler quotient of both theories in a bit more detail. First we begin with the $R_{0,4}$ side. While on a generic point the global symmetry is entirely broken, there are various subspaces where some symmetries are preserved. Particularly, there is a $1d$ subspace along which a $U(1)\times SU(4)^2$ subgroup is preserved. The symmetry breaking pattern is such that $SU(8)\rightarrow U(1)\times SU(4)^2$, and the preserved $U(1)$ is a combination of this $U(1)$ and the Cartan of the $SU(2)$. In the language of the class ${\cal S}$ description, along this subspace we are preserving the $SU(4)$ symmetries of the two maximal punctures, but break $U(1)_t$ completely and the $SU(2)\times U(1)$ symmetry of the next to maximal puncture is broken to a $U(1)$.

We can proceed and break the symmetry further by identifying the two $SU(4)$ groups and break the $U(1)$. This leads to a $4d$ subspace along which we preserve an $SU(4)$ global symmetry.

The conformal manifold on the quiver side is more complicated and we will not try to analyze it in detail. However, there appears to be a $4d$ subspace along which the  $U(1)$  groups of the free locus are broken, and one preserves only an $SU(4)$ embedded into the diagonal $SU(6)$ as its $SO(6)$ subgroup. We then find that both descriptions share a subspace with the same dimension and symmetry, and so it is possible that they can be related solely on that subspace without needing to break all the symmetries. We can check this proposal by comparing anomalies for these symmetries as well as the superconformal index refined by them.

 First we can consider matching the anomalies. In the $R_{0,4}$, since the theory has $\mathcal{N}=2$ supersymmetry, the only non-vanishing anomaly involving this symmetry is the $U(1)_R SU(4)^2$, which in turn is related to the contribution to the beta function upon gauging the symmetry. As a result it is sufficient to compare the latter. Since the $SU(4)$ comes from the diagonal of two maximal puncture, and each maximal puncture contributes half the matter required for an $\mathcal{N}=2$ $SU(4)$ gauge theory to be conformal, we get that gauging this $SU(4)$ contributes as $16$ fundamental chirals.

On the dual side the $SU(4)$ comes from the diagonal $SO(6)$ subgroup of the four $SU(6)$ groups, and therefore gauging it contributes as $8$ antisymmetrics of $SU(4)$. We first note that the antisymmetric is a real representation and so does not contribute to the cubic anomaly. This is good as that means that the $SU(4)^3$ anomaly matches. Finally, as the antisymmetric contributes like two fundamentals to the beta functions, we see that gauging the $SU(4)$ contributes as $16$ fundamental chirals, matching the result of the class ${\cal S}$ theory.    

We can next check the index. For this we can evaluate the index of the quiver theory refined by these symmetries. We find:
\be
{\cal I}=1+(6 + 4\; \bold{15}) (qp)^{\frac23}+(4 + 2\;\bold{20'}  + 2\; \bold{15}) qp+\cdots\,.\nonumber\\
\ee

This matches the index computed in \cite{Agarwal:2018ejn} when restricted to these symmetries. 

\

\subsection{Dual of ${\cal N}=2$ $T_4$ SCFT}

Next we consider  the $T_4$ ${\cal N}=2$ SCFT. This is a strongly coupled model which can be obtained in class ${\cal S}$ construction as a compactification of the $A_3$ $(2,0)$ theory on a sphere with three maximal punctures.  Several facts are known about this model. In particular the conformal anomalies are,
\be
a=19 \, a_v+99 a_\chi =\frac{45 }8\, , \; c= 19 \, c_v+99 c_\chi =\frac{13 }2\,.
\ee The model has $SU(4)^3$ global symmetry. It does not have dimension two Coulomb branch operators. However, it has a dimension three Coulomb branch operator as well as two dimension four ones that are not going to be important here. Additionally, it has dimension three Higgs branch chiral ring operators in the $(\bold{4},\bold{4},\bold{4})$ and $(\overline{\bold{4}},\overline{\bold{4}},\overline{\bold{4}})$ of $SU(4)^3$. This implies that there is an $\mathcal{N}=1$ only preserving conformal manifold. One can analyze the Kahler quotient and conclude that the conformal manifold is $83$ dimensional and on a generic point of which the $SU(4)^3$ global symmetry is completely broken. We also note for future use that this SCFT has $45$ relevant operators coming from the moment map operators with ${\cal N}=1$ R charge $\frac43$.

\begin{figure}[htbp]
\includegraphics[scale=0.62]{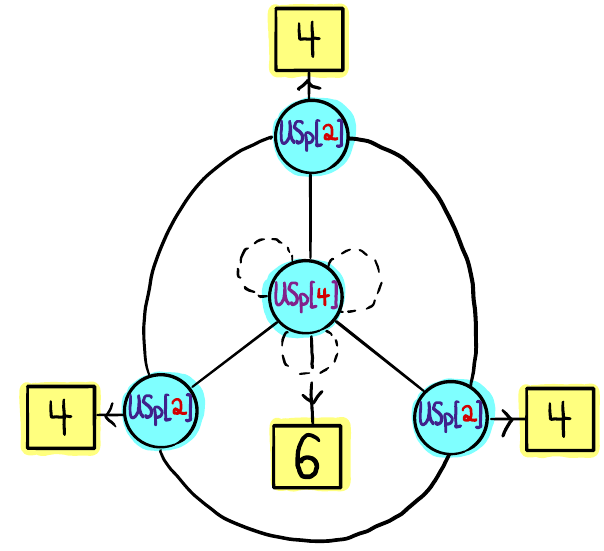}
\caption{The ${\cal N}=1$ conformal dual of $T_4$. The dashed lines are two index antisymmetric fields.}\label{F:T4}
\end{figure} 

As we have $19$ vector fields there are two possibilities for a conformal dual gauge group, $SU(3)^2\times SU(2)$ and $USp(4)\times SU(2)^3$. It turns out that it is possible to find a dual with the second option and it is depicted in Figure \ref{F:T4}. The number of chiral fields is $99$ and all the gauge groups are conformal.
The relevant operators are built from quadratics of bifundamentals and from the three two index antisymmetric fields, and a simple counting reveals that there are $45$ of those.

We can next compare the structure of the conformal manifold. In the quiver theory we have $178$ marginal operators. Four come from the bi-fundamental triangles, and $48$ come from the three gauge invariants going from one global $SU(4)$ to another. Nine come from the gauge invariants made from the $SU(2)\times USp(4)$ bifundamentals and the $USp(4)$ antisymmetrics, and $45$ come from the invariant made from the antisymmetric and two flavors of $USp(4)$. The rest come from the three gauge invariants going from one global $SU(4)$ to the global $SU(6)$. At the free point the theory has the non-anomalous $SU(6)\times SU(4)^3 \times SU(3)\times U(1)^7$ global symmetry. It is possible to show that there is a non-trivial Kahler quotient implying that this theory is conformal, and that on a generic point the global symmetry is broken completely. Therefore, this theory has an $83$ dimensional conformal manifold, matching the one of the $T_4$ SCFT.   

Finally, We can also match the superconformal index of the $T_4$ theory to the one computed using the gauge theory description. For example in the Schur limit $t=q$ \cite{Gadde:2011ik} the index was computed to high orders in expansion in $q$ in \cite{Beem:2017ooy} and it can be matched to the one computed using the Lagrangian.  As we tune the R charges to be free $t=(qp)^{\frac23}$ we need farther to set $q=p^2$ to obtain the Schur index.\footnote{We are grateful to Chris Beem and Carlo Meneghelli for pointing out this to us.}

\

\subsection{Dual of ${\cal N}=2$ $R_{2,5}$ SCFT}

We consider now  the $R_{2,5}$ ${\cal N}=2$ SCFT. This is a strongly coupled model which can be obtained in class ${\cal S}$ construction as a compactification of the $A_4$ $(2,0)$ theory on a sphere with one maximal puncture and two non-maximal punctures \cite{CD}\footnote{The punctures correspond to Young tableaux with a row with two boxes and row with three boxes.}. Several facts are known about this model. In particular the conformal anomalies are,
\be
a=14 \, a_v+86 a_\chi =\frac{53 }{12}\,, \;  c= 14 \, c_v+86 c_\chi =\frac{16 }3\,.
\ee The model has $SO(14)\times U(1)$ global symmetry. It does not have dimension two Coulomb branch operators but has one dimension three and one dimension five operators. It also has dimension three Higgs branch operators that are in the $\bold{64}$ and $\overline{\bold{64}}$ with equal but opposite charges under the $U(1)$. For convenience we shall normalize the $U(1)$ so that the Higgs branch operators have charge $\pm 1$. It is straightforward to see that these have a non-trivial Kahler quotient implying the existence of an ${\cal N}=1$ only preserving conformal manifold. One can work out the Kahler quotient and conclude that there is a $36$ dimensional conformal manifold on a generic point of which the symmetry is broken completely.  
We also note that it has $92$ relevant operators coming from the moment map operators with ${\cal N}=1$ R charge $\frac43$.

\begin{figure}[htbp]
\includegraphics[scale=0.62]{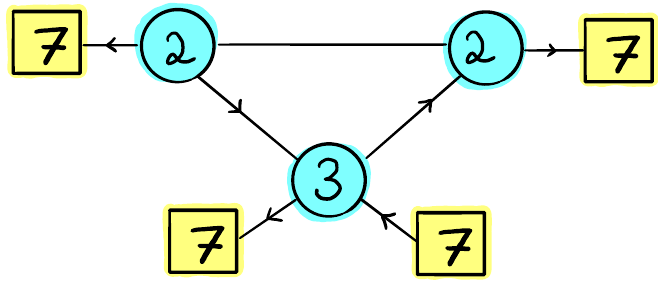}
\caption{The ${\cal N}=1$ conformal dual of $R_{2,5}$.}\label{F:R25}
\end{figure}

Like in the previous cases, we can seek a dual ${\cal N}=1$ gauge theory. Here, as in the $G_2$ gauge theory case, we have $14$ vectors and thus we should look for $SU(2)^2\times SU(3)$ gauge theory dual. The model which has all the right properties is depicted in Figure \ref{F:R25}. We have $86$ chiral fields and all gauge groups are conformal. The relevant operators are  mesons of $SU(3)$ which gives $49$ operators, quadratic combinations of bifundamentals of $SU(7)$ and $SU(2)$ giving $21+21=42$ operators,  and finally the quadratic combination of the bi-fundamental of the two $SU(2)$ gauge symmetries giving an additional operator. All in all we get $92$ operators as expected. 

We can next compare the structure of the conformal manifold for the two theories. We shall start with the quiver theory. Ignoring gauge coupling constants, we have $232$ marginal operators: one associated with the triangle, $147$ coming from the three flavor $SU(7)\times SU(7)$ bi-fundamental operators connected using each face of the triangles, $35+35$ $SU(3)$ baryons and anti-baryons from the fundamental flavors, and finally $7+7$ $SU(3)$ baryons and anti-baryons made from the bi-fundamental and one fundamental flavor. At the free point the gauge theory has $U(1)^4 \times SU(7)^4$ non-anomalous global symmetry. We can check that $232-4-4\times 48=36$ so the number of marginal operators minus conserved currents matches between both theories.

We next want to analyze the structure of the conformal manifold. First, it is straightforward to show that the triangle and the three $SU(7)\times SU(7)$ bi-fundamental operators have a Kahler quotient by themselves. Turning only them on then gives a $1d$ subspace along which the global symmetry is broken as $U(1)^4 \times SU(7)^4 \rightarrow U(1)\times SU(7)$. Here the preserved $SU(7)$ is a diagonal combination, which up to global charge conjugation can be chosen as $\bold{7}_{SU(2)_{left}} = \overline{\bold{7}}_{SU(2)_{right}} = \bold{7}_{SU(3)_{F}}= \overline{\bold{7}}_{SU(3)_{\overline{F}}}$. The preserved $U(1)$ is such that the $SU(3)$ fundamentals have charge $+4$, the antifundamentals charge $-4$, the $SU(3)\times SU(2)_{left}$ bi-fundamental has charge $+7$, the $SU(3)\times SU(2)_{right}$ bi-fundamental has charge $-7$, the $SU(2)_{left}$ flavors have charge $-3$, the $SU(2)_{right}$ flavors have charge $+3$, and the $SU(2)\times SU(2)$ bi-fundamental is uncharged.

At a generic point on this subspace the relevant and marginal operators take the form:
\be
&&{\cal I}_{rel} = (2 + \bold{48} + \frac{1}{x^6} \bold{21} + x^6 \overline{\bold{21}})(qp)^{\frac23}\, , \label{R25Rel}\\
&&{\cal I}_{mar} = (1 + x^{12} \bold{35} + \frac{1}{x^{12}} \overline{\bold{35}} + x^{18} \bold{7} + \frac{1}{x^{18}} \overline{\bold{7}} )(qp)\, , \label{R25Mar}
\ee
where we use the fugacity $x$ for the unbroken $U(1)$. 

We note two things about these expressions. First the marginal operators along this subspace form a self-conjugate representation. This suggests that we can indeed continue and break all the symmetry on a generic point on the conformal manifold, so that the conformal manifolds indeed agree between the two theories. Another interesting observation is that the relevant operators look like $1$ plus the adjoint of $SO(14)$ when represented using its $U(1)\times SU(7)$ subgroup. This suggests that the two theories may actually be related along this $1d$ subspace, and we shall next show that this conjecture passes several non-trivial tests.

First let us consider the conformal manifold of the $R_{2,5}$ SCFT. It also has a $1d$ subspace preserving $U(1)\times SU(7)$, where $SO(14)$ is broken to $U(1)\times SU(7)$, and the preserved $U(1)$ is a combination of this and the $U(1)$ part of the $R_{2,5}$ SCFT global symmetry. Specifically, consider using the fugacity $h$ for the $U(1)$ and decomposing $SO(14)$ to $U(1)_s\times SU(7)$ such that $\bold{14}\rightarrow \bold{7}s^2 + \frac{1}{s^2}\overline{\bold{7}}$, $\bold{64}\rightarrow \frac{1}{s^7} + \frac{1}{s^3} \bold{21} + s\overline{\bold{35}} + s^5\overline{\bold{7}}$. Then the preserved $U(1)$ is given by the identification $h=s^7$, corresponding to inserting the $SU(7)$ singlet opertaors appearing in the decomposition of the $\bold{64}$ and $\overline{\bold{64}}$ to the superpotential. It is straightforward to show that the relevant and marginal operators form the characters as in (\ref{R25Rel}) and (\ref{R25Mar}) if we identify $s^4 = \frac{1}{x^6}$.   

Finally we can compare anomalies. Since one side is an ${\cal N}=2$ SCFT, all anomalies for these symmetries must vanish save for the one linear in the $U(1)$ R-symmetry. Indeed in the quiver side it is easy to see that this is obeyed as the $SU(7)$ representations and $U(1)_x$ charges come in self-conjugate pairs. This leaves us with the $Tr(U(1)_R SU(7)^2)$ and $Tr(U(1)_R U(1)^2_x)$ anomalies. These can be computed to give:
\be
&&Tr(U(1)_R SU(7)^2) = -\frac{1}{3} \times \frac{1}{2} (2+2+3+3) = -\frac{5}{3},\\
&&Tr(U(1)_R U(1)^2_x) = -\frac{1}{3} (12 (7)^2 + 42 (4)^2 +28 (3)^2 ) = -504 .\nonumber
\ee
It is straightforward to show that the $SU(7)$ anomaly matches as the $SO(14)$ flavor central charge is such that the contribution to the beta function from gauging the $SO(14)$ is that of five chiral vectors. 

The $U(1)$ anomaly is trickier. To check it we utilize the duality in \cite{CDTA,CDT} where an ${\cal N}=2$ gauging of the $SO(7)$ subgroup of the above $SU(7)$ subgroup of the symmetry of $R_{2,5}$ SCFT is dual to an ${\cal N}=2$ $SU(6)$ gauge theory with one symmetric and one antisymmetric hypermultiplets. We can next compare the operators between the two dual sides. On the $SU(6)$ side we have two types of interesting marginal operators. One type is that of the Higgs branch operators built from baryons of the antisymmetric hyper, $\chi^3_{AS}$, $\chi^3_{\overline{AS}}$, while the other is that of the mixed branch operators $\chi_{AS} \Phi \chi_{\overline{S}}$, $\chi_{S} \Phi \chi_{\overline{AS}}$. These can be matched to the dual side. The Higgs branch operators must match to Higgs branch operators of the $R_{2,5}$ SCFT, and there are precisely two of these, the ones coming from the singlets in the decomposition of the spinors. The mixed branch operators come from the gauge invariant made from the $SO(7)$ adjoint chiral and the components in the moment maps in the adjoint representation of $SO(7)$. There are three of these with charges $s^4$, $0$ and $\frac{1}{s^4}$. The middle one is just the ${\cal N}=2$ preserving marginal deformation while the other two map to the mixed branch operators. Note that the mixed branch operators on one side then are not mapped to ones in the $R_{2,5}$. This is consistent with the duality as we have mapped the conformal manifold under the assumption that these are absent so their presence would spoil the duality.

We can now use the Lagrangian description to calculate the anomaly of the $R_{2,5}$ SCFT. As we previously mentioned, the marginal deformation that we are turning on is associated with the singlets in the spinor decomposition which in turn is mapped to the $SU(6)$ baryon from the antisymmetric. This breaks the antisymmetric $U(1)$, and the remaining $U(1)$, the one acting on the symmetric, should then map to the one preserved on the conformal manifold. We can now compute its anomaly from the $SU(6)$ theory:
\be
Tr(U(1)_R U(1)^2_x) = -\frac{1}{3} 42 (6)^2= -504 ,
\ee         
and the anomaly non-trivially matches.

Note that this implies that the $SU(6)+1S+1AS$ gauge theory has a Lagrangian dual given by the $SO(7)$ gauging of all the $SU(7)$ groups in the quiver theory, which should emerge when going on a $1d$ subspace of its conformal manifold associated with turning on the operators $\chi^3_{AS}$, $\chi^3_{\overline{AS}}$. As both sides are Lagrangian we can compare the full indices of both theories, refined under the preserved $U(1)$, which is a non trivial check of this proposal. Furthermore, the $R_{2,5}$ SCFT participates in another duality \cite{CD} where gauging the $USp(4)$ group in the $SU(2)\times USp(4)\times SO(6)$ subgroup of $SO(14)$ with one fundamental hyper for the $USp(4)$ is dual to the ${\cal N}=2$ gauge theory $SU(5)+2AS+4F$\footnote{The ${\bf 7}_{SU(7)}$ becomes $3\times {\bf 1}_{USp(4)}\oplus{\bf 4}_{USp(4)}$.}.  We can use this as well to also get an ${\cal N}=1$ gauge theory dual for $SU(5)+2AS+4F$, and for another consistency check.   We have compared indices in expansion in fugacities and verified that to order $(qp)^{\frac43}$ in expansions the indices indeed agree giving additional support for our conjecture. Note that these two dualities are between ${\cal N}=2$ conformal Lagrangian theories and conformal Lagrangians having only explicitly ${\cal N}=1$ supersymmetry.

\

\subsection{Dual of ${\cal N}=2$ rank $2n$ $E_6$ SCFT}

As our final class ${\cal S}$ example we consider the case of the Minahan-Nemeschansky $E_6$ theories for general rank. Among the three Minahan-Nemeschansky $E$ type theories this is the only family with a dimension three Coulomb branch operator. Additionally, for rank $2$ and higher, it also has dimension three Higgs branch operators, and as a result can have an ${\cal N}=1$ preserving conformal manifold. This class of theories have a rather uniform behavior at high rank. Specifically, for rank higher than $1$ they have an $SU(2)\times E_6$ global symmetry and a dimension three Higgs branch operator in the $(\bold{2}, \bold{78})$. Above rank $2$ they also have a dimension three Higgs branch operator in the $\bold{4}$ of the $SU(2)$. As a result it is not difficult to work out the Kahler quotient for the entire family. Specifically, we find that when the rank is $2$ there is a $75$ dimensional conformal manifold on a generic point of which the global symmetry is completely broken. This remains also for the higher rank cases, but the dimension of the conformal manifold grows to $79$ due to the existence of the additional marginal operators.

We again note that this theory has $81$ relevant operators associated with the $SU(2)\times E_6$ moment maps.

\begin{figure}[htbp]
\includegraphics[scale=0.62]{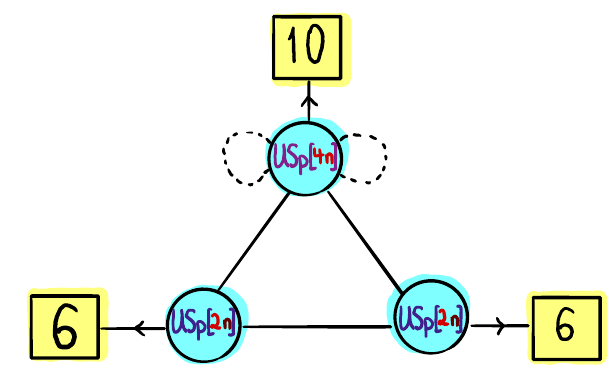}
\caption{The ${\cal N}=1$ conformal dual of rank $2n$ $E_6$ SCFT. The dashed lines are two index antisymmetric fields.}\label{F:E6rankQ}
\end{figure} 
Like the previous cases we seek an ${\cal N}=1$ Lagrangian dual on the conformal manifold.  The conformal anomalies of the rank $N$ $E_6$ SCFT give \cite{Aharony:2007dj},
\be
&&a=n_v(N) \, a_v+n_\chi(N) a_\chi =\frac34 N^2+N-\frac1{24}\, ,\\
&& c= n_v(N) \, c_v+n_\chi(N) c_\chi =\frac{3 }4 N^2+\frac32 N-\frac1{12}\,, \nonumber
\ee with
\be 
n_v(N)=N(3N+2)\,,\;\; n_\chi(N)=9N^2+30N -2\,.
\ee 
For even $N=2n$ the quiver in Figure \ref{F:E6rankQ} has the right number of fields and the gauge groups are conformal. We have a quiver gauge theory with two $USp(2n)$ and one $USp(4n)$ groups.  We conjecture that this quiver is dual to the rank $2n$ $E_6$ Minahan-Nemeschansky model. We can first count the relevant operators. 
These are given by antisymmetric squared combinations of fundamentals, and by symmetric squared of the antisymmetric and bifundamentals fields. All in all we get $2\frac{6\times 5}2+\frac{10\times 9}2+3+3=81$  as expected. Next we can analyze the conformal manifold. The non-anomalous symmetry at the free point is $SU(6)^2\times SU(10)\times SU(2)\times U(1)^4$, giving $176$ currents. The computation of the number of marginal deformations differes between $n=2$ and higher $n$. For $n>2$ the cubic symmetric power of the antisymmetric fields contains a singlet and it does not for $n=2$. Thus these contribute additional $\frac{2\times 3\times 4}6=4$ operators for $n>2$ as expected.  The rest of the marginal operators are $60+60+36+1+2\times \frac{10\times 9}2+4=251$. Assuming that all the symmetry is broken on general point of the conformal manifold we get the expected result of $75$ deformations for $n=2$ and $79$ for $n>2$.

\section{Conformal duals of ${\cal N}=1$ $E_8$ SCFTs}

One can obtain strongly coupled SCFTs by taking arbitrary $(1,0)$ theories in six dimensions and compactifying them on Riemann surface with a proper twist to preserve ${\cal N}=1$ supersymmetry. In general these models are some SCFTs for which Lagrangian constructions are not known. Let us discuss here one example of such an interesting model and construct a conformal Lagrangian for it. 
We consider the $6d$ rank one E-string theory compactified  on genus $g$ Riemann surface with zero flux for its global $E_8$ symmetry. One can compute anomalies of this model \cite{Kim:2017toz} and find that for $g>1$,
\be
\frac{a}{g-1}=16 a_v+81 a_\chi=\frac{75}{16}\,,\, \frac{c}{g-1}=16 c_v+81 c_\chi=\frac{43}{8}\,.\nonumber\\
\ee The number of vectors is thus $16(g-1)$.
This implies for example that for genus two we might have a conformal description with two $SU(3)$ groups or $USp(4)$ and two $SU(2)$ groups,  and $81$ chiral fields. We find a dual using two $SU(3)$ groups.
Several facts are known about this model. First it has no supersymmetric relevant deformations and second it has a conformal manifold on which the $E_8$ global symmetry is completely broken on a generic locus. The dimension of the manifold is,
$dim {\cal M}_c=3+248+1$. Here  $3=3g-3$ comes from complex structure moduli, $248=dim E_8 (g-1)$ comes from flat connections, and the additional $1$ is a deformation which does not have a generic origin as the rest but as the index of this theory is known \cite{Kim:2017toz} it can be inferred from it using \cite{Beem:2012yn}.
\begin{figure}[htbp]
\includegraphics[scale=0.62]{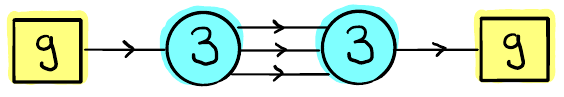}
\caption{The  conformal dual of rank one E-string compactified on genus two surface with no flux.}\label{F:Estringd}
\end{figure} 

We then can construct a putative dual depicted in Figure \ref{F:Estringd}. The model has no supersymmetric relevant deformations as one cannot build mesonic operators. The marginal operators come from baryons, $2(\frac{9\times 8\times 7}6)+10$, and from cubic composites winding the quiver, $9\times 3\times 9$. All in all we have $421$ marginal operators. The non-anomalous symmetry at the free point is $SU(9)^2\times SU(3)\times U(1)$ which gives us $80+80+8+1=169$ currents. On a general point of the conformal manifold all the symmetry is broken and we obtain $252$ exactly marginal operators as expected. We also can compute the index and compare it to the one obtained in \cite{Kim:2017toz} and find a match to low orders in expansion in fugacities. We thus conjecture that this is a Lagrangian description of the compactification of rank one E-string on genus two surface with no flux. As a result, we expect that on some point on the conformal manifold of this model the symmetry is expected to enhance to $E_8$ \footnote{ Incidentally the same quiver has a different geometric origin as twisted compactification on a sphere with four maximal twisted punctures of minimal $SU(3)$ $(1,0)$ SCFT \cite{Razamat:2018gro}. Such equivalences of compactifications are common but deep understanding of them is lacking at the moment.}.
\begin{figure}[htbp]
\includegraphics[scale=0.52]{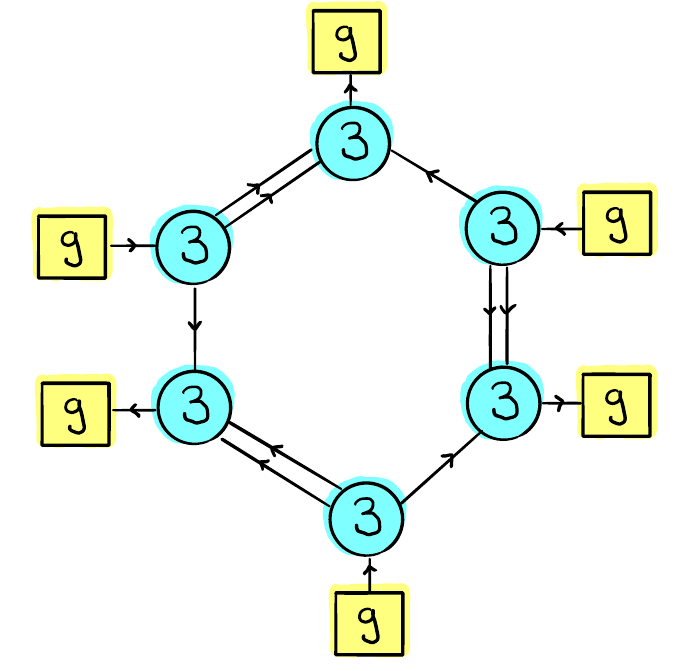}
\caption{The  conformal dual of rank one E-string compactified on genus $g$ surface with no flux. The number of gauge nodes is $2g-2$. Here we have example of $g=4$.}\label{F:estringdg}
\end{figure} 

We can generalize the above to arbitrary genus $g>2$. The quiver theory is in Figure \ref{F:estringdg}. The number of fields is as expected. We have $8\times (2g-2)=16(g-1)$ vectors and $27\times (2g-2)+9\times (3g-3)=81(g-1)$ chiral fields. We have no relevant operators. The marginals are $(4+1)(g-1)+84 (2g-2)$ baryons, and $81\times (2+1)\times (g-1)$ cubic combinations winding between the $SU(9)$ groups. All in all this gives $416(g-1)$ marginal operators. The symmetry at the free point is $SU(9)^{2g-2}\times SU(2)^{g-1}\times U(1)^{2g-2}$ giving $165(g-1)$ currents. The symmetry is broken on a general point of the conformal manifold and thus we have $251(g-1)=3g-3+248(g-1)$ exactly marginal deformations as expected. Note that when $g=2$ an $SU(2)\times U(1)$ symmetry enhances to $SU(3)$ giving us $8$ instead of $4$ currents, and we have $10$ baryons instead of $4+1$ for bifundamentals of $SU(3)$. Thus, relative to the general case we have four more currents and five more marginal operators giving us an additional exactly marginal direction. Thus we conjecture that the quiver of Figure \ref{F:estringdg} describes the compactification of rank one E-string on genus $g$ surface with zero flux. In particular we expect the symmetry to enhance to $E_8$ somewhere on the conformal manifold.
\begin{figure}[htbp]
\includegraphics[scale=0.52]{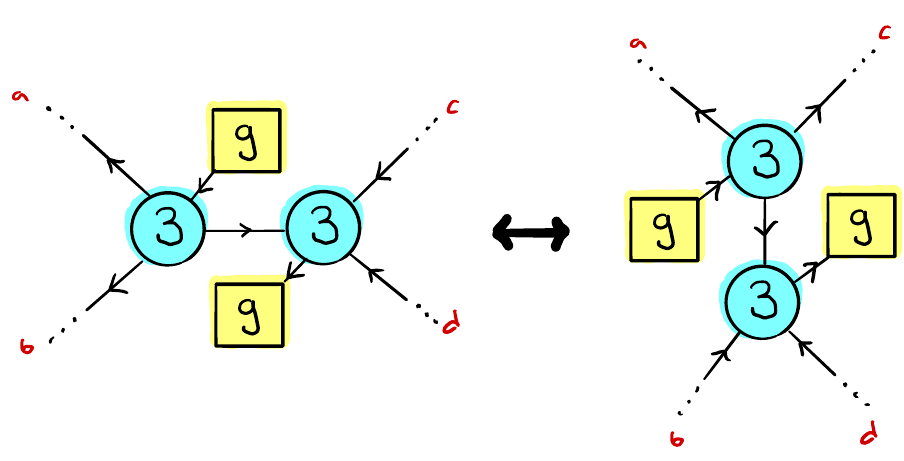}
\caption{The index of a genus $g$ quiver is consistent with such crossing symmetry duality moves. We assume that superpotentials are turned on breaking all the symmetry.}\label{F:dualitymove}
\end{figure}

Note that we can turn on superpotentials which break only part of the symmetry. In particular turning on cubic superpotentials locking all the $SU(9)$ symmetries together, as well as baryonic superpotentials for the bifundamentals, we obtain a submanifold of ${\cal M}_c$. Moreover $E_8$ has an $SU(9)$ maximal subgroup with the decomposition ${\bf 248}_{E_8}\to {\bf \overline {84}}\oplus {\bf 84} \oplus {\bf 80}$. We note that all the representations of $SU(9)$ in the index combine into $E_8$ representations under this branching rule. For example, for the theory corresponding to genus $g$, after identifying all $SU(9)$ groups, it is easy to see that we get $(g-1){\bf 248}_{E_8}$ exactly marginal operators such that $g-1$ ${\bf 84}$ and ${\bf \overline{84}}$ come from baryons and $g-1$ ${\bf 80}$ coming from operators winding between $SU(9)$ groups.  We have more operators which are singlets of $SU(9)$ whose number is $3g-3$. Furthermore, the $Tr(U(1)_R SU(9)^2)= -(g-1)$ matches the result expected from the strongly coupled theory where the $SU(9)$ is embedded inside $E_8$. All this suggests that the two theories might both sit on a shared subspaces where at least the Cartan subalgebra of $E_8$ is preserved.
\begin{figure}[htbp]
\includegraphics[scale=0.42]{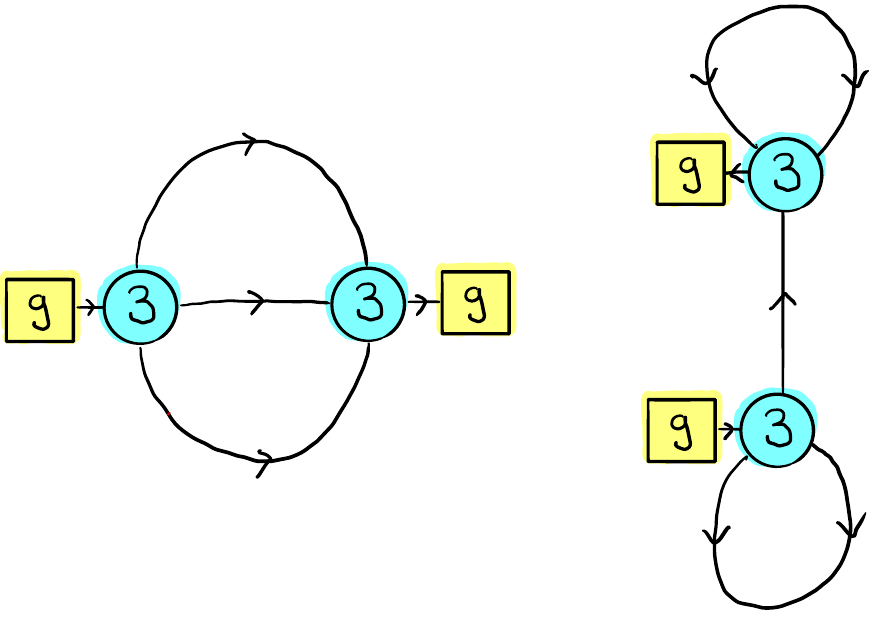}
\caption{The two different pair-of-pants decompositions of the genus two theory corresponding to Figure 9 and Figure 13.}\label{F:dualitymoveg2}
\end{figure} 
 
\begin{figure}[htbp]
\includegraphics[scale=0.62]{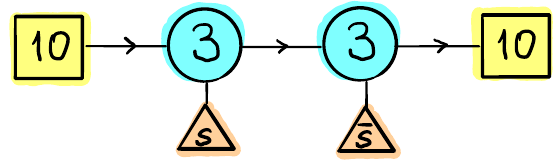}
\caption{The crossing symmetry move implies this dual for genus two theory.}\label{F:dualgenus2}
\end{figure}

Note that in the dual of genus $g$ compactification of E-string the number of gauge groups is $2g-2$ and number of bifundamentals is $3g-3$ suggesting  a possibility of  a  geometric interpretation with gauge groups playing the role of the ``pairs of pants'' and the bi-fundamental fields being the tubes. The pairs-of-pants then are associated to the $SU(3)$ gauge groups and the tubes to the bi-fundamentals. This interpretation is different than the usual class ${\cal S}$ logic where pairs-of-pants are matter and tubes are gauge groups. If such a geometric interpretation is correct we expect that the theories will be invariant under duality moves a la crossing symmetry, see Figure \ref{F:dualitymove}. We indeed find that with generic superpotentials such duality moves give equivalent theories as far as anomalies and indices are considered. In particular such a duality move for genus two theory, Figure \ref{F:Estringd}, gives the model in Figure \ref{F:dualgenus2}. Here we ``glue'' two punctures of the same pair-of-pants together, see Figure \ref{F:dualitymoveg2}. The tubes are associated to the bi-fundamental matter and here the group is the same meaning ${\bf 3}\times {\bf 3}\to {\bf 6}\oplus {\bf \overline 3}$. We checked for the index that indeed this duality holds. 

\

\section{Discussion} In this paper we have discussed a simple algorithm to seek for conformal gauge theory duals of SCFTs. We have illustrated the power of this simple procedure by finding the duals of a variety of interesting models. Some of these models are strongly coupled SCFTs and our procedure provides a conformal Lagrangian description for them. Although we concentrated on specific examples the procedure we have outlined here can be used to systematically search for conformal gauge theory duals of any SCFT.  For example we can extend the discussion to duals of theories with non simple gauge groups, or in class ${\cal S}$ setting to theories corresponding to more complicated surfaces. The procedure can be modified for the sought after dual to include also strongly coupled ingredients. 

Our procedure has provided conjectured dual Lagrangians for some of the strongly coupled SCFTs so let us make several comments on this. First, some of the strongly coupled SCFTs have already Lagrangian constructions of two types. First, one can construct ``singular Lagrangians'' by starting with a Lagrangian gauge theory $T$ and gauging a symmetry which appears only at some strong coupling cusp \cite{Gadde:2015xta}. This procedure was applied for rank one $E_6$ MN model in \cite{Gadde:2015xta}, to $R_{0,4}$ in \cite{Agarwal:2018ejn}, to some of the class ${\cal S}_k$ theories \cite{Gaiotto:2015usa} in \cite{Razamat:2016dpl}, and to some of the theories engineered by the compactification of the $6d$ rank one E-string SCFT \cite{Kim:2017toz}. Second type of Lagrangians was obtained by engineering the models of interest, such as Argyres-Douglas theories, as  IR fixed points of an RG flow \cite{Maruyoshi:2016tqk}. All these constructions have an RG flow and the symmetries and/or supersymmetries of the model of interest are not manifest in the description. However, the symmetry seen in the UV is of the same rank as the symmetry of the fixed point. The descriptions we have found do not have an RG flow on one hand but on the other hand the rank of the symmetry of the Lagrangian description on a generic point on the conformal manifold is smaller than the rank of the global symmetry of the dual strongly coupled SCFT. In particular, consider constructing a model $T$ by gauging some symmetry of two theories, $T_1$ and $T_2$, for which a conformal gauge theory dual has been found using our procedure. Then if for either $T_1$ or $T_2$ the conformal gauge theory dual cannot be reached by going on the conformal manifold without breaking the gauged symmetry then the conformal dual of $T$, if exists, is a priori not obviously related to those for $T_1$ and $T_2$. It would be very interesting to figure out if an interesting relation exists.

Let us also mention that there are other examples of conformal ${\cal N}=1$ dualities discovered in recent years which have a geometrical interpretation by engineering the models of interest as compactifications of six dimensional SCFTs. For example the duality group acting on the conformal manifold of $SU(3)$ SQCD theory with nine flavors was argued to be related to the mapping class group of a ten punctured sphere in \cite{Razamat:2018gro}, and conformal dualities between intricate quiver gauge theories following from five dimensional dualities were deduced in \cite{Kim:2018bpg}. It will be also extremely interesting to understand whether the dualities suggested here have a geometrical interpretation of a sort. 

Of course it will be extremely interesting to understand whether the conformal gauge theory duals obtained in the procedure discussed here follow any interesting patterns and satisfy some general rules.

\

\noindent{\bf Acknowledgments}:~
We would like to thank   Chris Beem, Leonardo Rastelli, and Yuji Tachikawa for very useful conversations. SSR would like to thank the organizers of the Pollica summer workshop for hospitality during final stages of this project. The workshop was funded in part by the Simons Foundation (Simons Collaboration on the Non-perturbative Bootstrap) and in part by the INFN, and the authors are grateful for this support. The research of SSR was  supported by Israel Science Foundation under grant no. 2289/18 and by I-CORE  Program of the Planning and Budgeting Committee.
 GZ is supported in part by  World Premier International Research Center Initiative (WPI), MEXT, Japan.

\end{document}